\documentclass[epj]{svjour}
\usepackage{graphics}
\begin{document}
\title{A method to polarise antiprotons in storage rings and
    create polarised antineutrons
     }
\author{Berthold Schoch
}                     
%
%
\institute{Physikalisches Institut, Universit\"at, Bonn, D-53115 Bonn, Germany }
\date{Received: date / Revised version: date}
%
\abstract{An intense circularely polarised $\gamma$-beam interacts with a cooled antiproton
beam in a storage ring. Due to spin dependent absorption cross sections for
the reaction $\gamma+\overline{p}\rightarrow \pi ^{-}+\overline{n}$ a
built-up of polarisation of the stored antiprotons takes place. Figures-of-merit around 
0.1 can be reached in principle over a wide range of antiproton energies. In this process polarised
antineutrons with polarisation $ P_{\overline{n}}\succ 70\%$ emerge.
The method is presented for the case of a 300 MeV/c
cooled antiproton beam.
  }

\PACS{
      { 13.88.+e}{polarisation in interactions and scattering}   \and
      { 29.20.Dh}{Storage rings}   \and
	  { 29.27.Hj}{polarised beams}
     } 
%
\maketitle
\section{Introduction}
\label{sec:1}
   The preparation of polarised beams of nucleons and antinucleons at energies in the GeV region poses
   challenging technical problems. In the case of protons powerful sources of polarised protons exist.
   Starting from low energies a cascade of accelerators, all equipped with
   polarisation conserving beam optics, brings the proton beam to its design energy.
   Antiprotons, however, are produced with high energy proton beams. Large acceptance storage rings are
   used to collect and cool the produced antiprotons in order to prepare an antiproton beam for physics experiments.
   The exploitation of the polarisation degrees of freedom in physics experiments
    plays for antiproton and antineutron induced 
   reactions a similar 
     important role as for other probes \textit{e.g.} electrons and protons. Challenging physics questions can be 
	addressed with polarised antiproton beams as soon as moderate
   luminosities, compared to polarised proton beams, can be achieved.
   In recent publications \cite{Ref1,Ref2,Ref3} the physics motivations and foreseen experimental 
   programs are addressed,
	previous studies to prepare a  polarised antiproton beam  are discussed and a  method 
	for the preparation of a polarised antiproton beam in
	the GeV region has been proposed \cite{Ref1}. In that proposal antiprotons would be polarised by the spin dependent 
	interaction in an
	electron polarised hydrogen gas target. 
	The antiproton beam polarisation $P_{\overline{p}}$ would reach $P_{\overline{p}}$=0.2-0.4.
      
The method described in this paper makes use of different reaction cross sections for the
two spin projections
of antiprotons stored in a storage ring interacting  with polarised $\gamma$-radiation.
The main ingredients driving the method are described. Limits and 
possible future directions of the application of the method are
addressed. 
Recent developments in accelerator physics   
paved the way to the proposed scheme.
In sect. 2 the method will be described. In sect. 3 the absorption cross sections will be presented . 
Sect. 4 deals with practical considerations for an implementation 
of the method. Sect. 5 addresses the question of preparing an antineutron beam. 
The article ends with a summary in sect. 6.  
\section{Method}
\label{sec:2}
The method makes use of the different sizes of total absorption cross sections for circularly
polarised photons on the antiproton. The antiprotons stored in a ring equipped
with a Sibirian snake \cite{Ref4} traverse a straight section.
\begin{figure}{}
\resizebox{0.45\textwidth}{!}{%
  \includegraphics{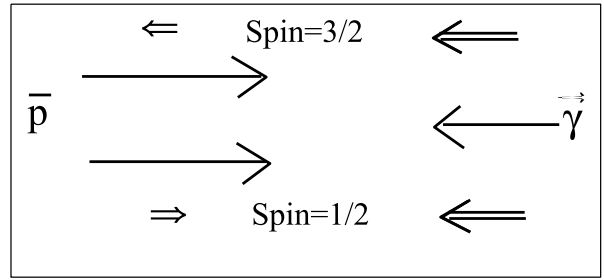}
}
\caption{The interaction of circularly polarised photons
with stored antiprotons.}
\label{fig:1}       
\end{figure} 
By shining circularly polarised photons against a stream of antiprotons,
see fig. 1, total absorption of the photons occurs, provided the photon
energy $E_{\gamma }$ exceeds the pion production thresholds of the
reactions $\gamma +\overline{p}\rightarrow \pi ^{-}+\overline{n}$, with $%
E_{threshold}^{\pi ^{-}}=151.43$ MeV, and $\gamma +\overline{p}\rightarrow
\pi ^{0}+\overline{p}$, with $E_{threshold}^{\pi ^{0}}=144.68$ MeV. Due to
the two possible total spin states of j=3/2 and j=1/2 two total absorption
cross sections, $\sigma _{3/2}$ and $\sigma _{1/2}$, come into play. The
ideal situation  of the application of the method would be reached when 
one of those cross sections would be
zero. Then the non disappearance of the other cross section would lead to
the loss of that component of the beam in the ring, provided that the
absorption process transforms the antiproton into an antineutron or, for the 
$\gamma +\overline{p}\rightarrow \pi ^{0}+\overline{p}$ -reaction, the
antiproton in the final state escapes the acceptance of the storage ring.

\subsection{Cross sections, luminosity}
\label{sec:3}
Fig. 2 shows those total cross sections $\sigma _{1/2}$ and $\sigma _{3/2}$
as a function of the photon energy $E_{\gamma }$ up to 500 MeV.
\begin{figure}{}
\resizebox{0.54\textwidth}{!}{%
  \includegraphics{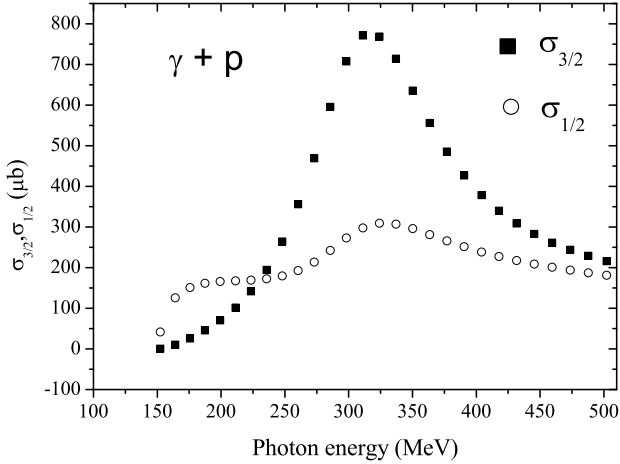}
}
\caption{The total absorption cross sections $\sigma _{1/2}$ and
$\sigma _{3/2}$ calculated with the MAID program \cite{Ref5}. }
\label{fig:2}       
\end{figure}
Two 
energy windows can be seen that can be used to reach high polarisation. A beam 
polarisation
\begin{equation} 
P(t)_{\overline{p}}=\frac{N(t)_{1/2}^{\overline{p}%
}-N(t)_{3/2}^{\overline{p}}}{N(t)_{1/2}^{\overline{p}}+N(t)_{3/2}^{\overline{p}}}
\end{equation}
develops during the interaction period, with $N(t)_{1/2}^{\overline{p}}$ 
and $N(t)_{3/2}^{\overline{p}}$ 
as the stored antiprotons of the 1/2 and 3/2 channels, respectively.
\vspace{1pt}Close to the $\pi ^{+}$ and $\pi ^{0}$ production thresholds,
around $E_{\gamma }=170 MeV,$ a cross section ratio of 9 can be reached,
thus, approaching almost the ideal situation. On top of the first antinucleon
resonance $\overline{\Delta }$, a second window for an efficient beam
polarisation opens up, a little bit wider on the energy scale than the first
one yielding a cross section ratio $\frac{\sigma _{3/2}}{\sigma _{1/2}}$ of
2-3.
Reaction rates yield "decay constants" of
the two components of $a_{1/2}$  and $a_{3/2}$.  
As an example that sets the goal for the application of the method, values of
$a_{3/2}=$ $3.6\cdot 10^{-5}$ $s^{-1}$ and $a_{1/2}$=$1.4\cdot 10^{-4}$ $s^{-1}$ have been chosen. For those values
of the decay constants and using a storage ring filled with  $N_{\overline{p}}=\dot 10^{10}$
unpolarised antiprotons the number 
of the stored antiproton develops in time as
\begin{equation}
N_{\overline{p}}=N_{1/2}^{\overline{p}}+N_{3/2}^{\overline{p}}
\end{equation}
\begin{equation}
N_{\overline{p}}=0.5\cdot 10^{10}\cdot (e^{-1.4\cdot 10^{-4}\cdot t}+e^{-3.6\cdot
10^{-5}\cdot t})
\end{equation}
That development of $N_{\overline{p}}$,  the built-up of the
polarisation $P_{\overline{p}}$ and figure of merit $FOM$ $=N_{\overline{p}%
}(t)\cdot (P_{\overline{p}}(t))^{2}$ are shown in fig. 3.
\begin{figure}{}
\resizebox{0.42\textwidth}{!}{%
  \includegraphics{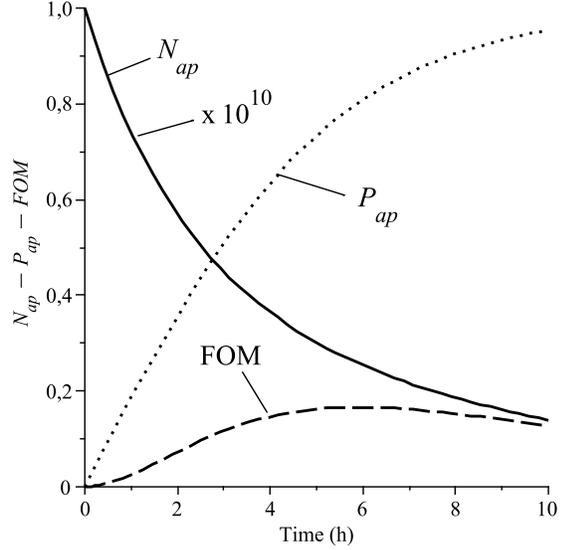}
}
\caption{Built-up of polarisation $P_{ap}$ and figure-of-merit (FOM)  as well
as a decreasing number of antiprotons in a storage ring.
}
\label{fig:3}       
\end{figure}
At the maximum of the figure-of-merit with $ FOM=0.18$ the number
of stored antiprotons decreases to $N_{\overline{p}}\approx  3\cdot 10^{9}$ with a polarisation of
$P_{\overline{p}} \approx 0.62$. In the ideal case, with only
one absorption channel present, the figure-of-merit could reach a value
of $FOM=0.5.$
Using the value of the decay constant $a_{1/2}$ and taking the cross section, \textit{e.g.} 165 $ \mu b$ at
$E_{\gamma} = $190 MeV, shown in fig. 2, the luminosity L
results to:
\begin{equation}
 L = 4.2\cdot 10^{33}cm^{-2}s^{-1}
 \end{equation}     
The question will be, whether experimental conditions can be prepared that
allow to approach that described goal as shown in fig. 3. 
Thus, one of the most important tasks
consists in preparing an intense circularly polarised photon beam.

\subsection{Energy, intensity and polarisation of a photon beam}
\label{sec:4}
\subsubsection{Energy of the photon beam}
\label{sec:5}
The method considered could possibly cover a large range of 
antiproton energies up into the GeV/c region.
In order to define the range of energies
two antiproton momenta are considered, 10 GeV/c and 300 MeV/c.
For an antiproton beam with a
momentum of 10 GeV/c the velocity in units of the velocity of light is
$\beta =\allowbreak$ 0.9956 and  $\gamma=\sqrt{\frac{1}{1-\beta ^{2}}}=\allowbreak 10.\,\allowbreak 704\,6$.
The energy region close the pion production threshold, $150\leq E_{\gamma }^{%
\overline{p}-system}/MeV \leq 200$ is considered first.
The photon energy in the laboratory system for 200 MeV\vspace{1pt} photons in the
rest system of the antiprotons amounts to
$E_{\gamma }^{lab}=\frac{E_{\gamma }^{\overline{p}-system}}{\gamma \cdot
(1+\beta )}=\allowbreak
9.\,\allowbreak 36$ $MeV$.
Working in the first resonance region with photons of 310 MeV needs a photon
source in the laboratory of
$E_{\gamma }^{lab}
=\allowbreak 14.\,\allowbreak 58$ $MeV$.
The respective values  are $E_{\gamma }^{lab}=146.03$ MeV for $E_{\gamma }^{%
\overline{p}-system}=200$ MeV and $E_{\gamma }^{lab}=226.43$ MeV for $%
E_{\gamma }^{\overline{p}-system}=310$ MeV 
for the case of antiproton momenta of 300 MeV/c in a storage
ring. The method will be discussed for the 300 MeV/c case.

\subsubsection{A suitable photon source}
\label{sec:6}
Electron bremsstrahlung provides the only powerful source for that
$\gamma$-energy range.
In the bremsstrahlung process, a helicity transfer from the electron to
the photon takes place. An almost completely circularly polarised
photon beam with the polarisation $P_{circ}$, close to the endpoint of the
photon spectrum, can be prepared by a longitudinally polarised electron beam.
Fig. 4 shows the degree of helicity transfer h \cite{Ref6} 
for a beam of bremsstrahlung with a 100\% polarised electron beam as a
function of $\varepsilon =\frac{E_{\gamma }}{E_{e}}.$\begin{equation}
h=\frac{
P_{circ}}{P_{e}}:=\varepsilon \frac{\frac{4}{3}-\frac{1}{3}\varepsilon }{%
\frac{1}{3}+\frac{2}{3}\varepsilon +\left( 1-\varepsilon \right) ^{2}}
\end{equation}
\begin{figure}{}
\resizebox{0.35\textwidth}{!}{
  \includegraphics{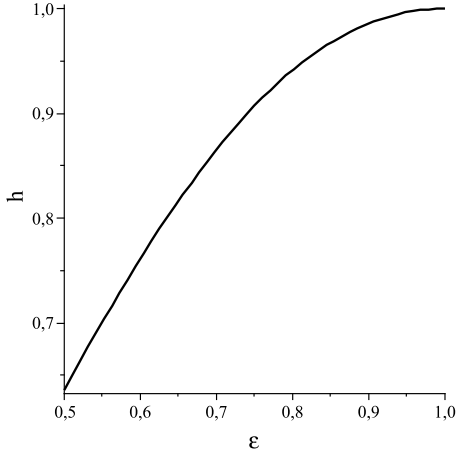}
}
\caption{Helicity transfer h as a function of $\varepsilon =\frac{E_{\gamma }}{E_{e}}$.
}
\label{fig:4}       
\end{figure} 

\section{A closer look to the absorption cross sections and polarised
electron sources}
\label{sec:7}
The applicability of the method depends on the knowledge of the input data,
 such as total absorption cross sections, $%
\sigma _{1/2}$ and $\sigma _{3/2}$.  
However, other issues will be decisive for
efficient applications such as high
luminosities. In addition, each
implementation of such a method depends on the beam optics of the storage
ring for a beam of polarised antiprotons and needs a cooled
beam. Each energy regime needs special considerations. A new facility FAIR \cite{Ref7}
will be built at GSI in Darmstadt, Germany.
A storage ring, HESR, will be able to store $10^{10}-10^{11}$ antiprotons in
the energy regime $3 \preceq E_{\overline{p}}/GeV \preceq 14$. It is anticipated that 
$N_{\overline{p}}=10^{10}$\textit{\ }is \textit{\ }the number of antiprotons
produced at FAIR in 15 min.
$N_{\overline{p}}=10^{10}$ will be also the number of antiprotons in the low
energy ring 
considered in the following discussion.

\subsection{Experimental data for the total photon proton cross sections}
\label{sec:8}

In the last few years the  cross section difference $(\sigma
_{3/2}-\sigma _{1/2})$  on the proton was measured in the photon energy range 
$200\preceq E_{\gamma }/MeV\preceq 2800$ at the accelerators MAMI, Mainz \cite{Ref8},
and ELSA, Bonn \cite{Ref9}.
From the total absorption cross section $\sigma _{tot}$, compiled by the particle data group \cite{Ref10},
and the cross section difference
$(\sigma _{3/2}-\sigma _{1/2})$ the cross sections for circularly polarised
photons, $\sigma _{3/2}$ and $\sigma _{1/2}$, can be extracted:
\begin{equation}
\sigma _{3/2}=\sigma _{tot}+\frac{(\sigma _{3/2}-\sigma _{1/2})}{2}
\end{equation}
\begin{equation}
\sigma _{1/2}=\sigma _{tot}-\frac{(\sigma _{3/2}-\sigma _{1/2})}{2}
\end{equation}
\newline
\begin{figure}{}
\resizebox{0.45\textwidth}{!}{%
  \includegraphics{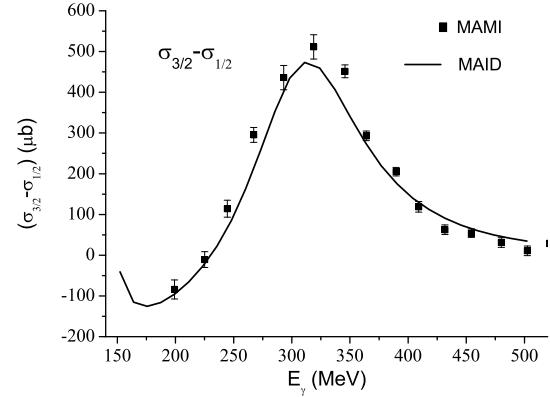}
}
\caption{Data and results of calculations
 with the MAID program for the difference
$\sigma _{3/2}-\sigma _{1/2}$ .
}
\label{fig:5}       
\end{figure}
\begin{figure}{}
\resizebox{0.45\textwidth}{!}{%
  \includegraphics{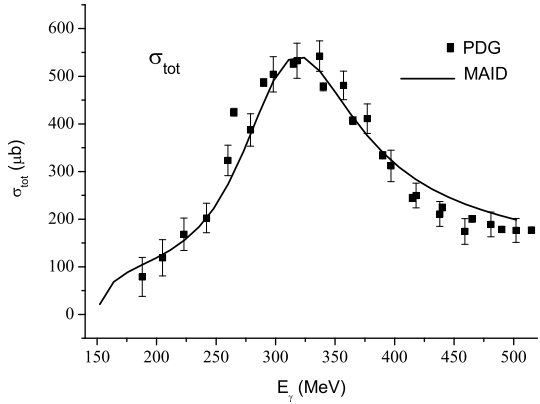}
}
\caption{Data and results of calculations
 with the MAID program for the total absorption
 cross section.
}
\label{fig:6}       
\end{figure}
Photoproduction data have been analyzed and described
by the results of calculations based on a phenomenological model.
The results of calculations with the computer
program MAID \cite{Ref3} are shown together with the data in fig(s). 5 and 6.
For the purpose of this paper the agreement between the MAID results and the
measurements are good enough in order to use the MAID results for the
polarised photon cross sections $\sigma _{1/2}$ and $\sigma _{3/2}$, see
fig. 2. The MAID results allow also to cover the near pion threshold region,
not covered by the mentioned experiments.
\subsection{Polarised electrons}
\label{sec:9}
Polarised electrons are routinely produced via photo emission
from GaAs-crystals either in pulsed  or dc mode  
by LASER light, \textit{e.g.} ref. \cite{Ref11} and ref. \cite{Ref12}, respectively. 
Average currents of $%
I_{e}=200$ $\mu A$ with a beam polarisation of $P_{e}=0.85$ are achieved.
The emittances reached are well adapted to use them for interactions with cooled
hadron beams. On several places research is going on to increase the current.
At Jefferson Lab. \textit{e.g.} a dc polarized electron beam with a current of 1 mA 
has been produced recently.

\section{Practical considerations}
\label{sec:10}
\subsection{A possible implementation of a set-up}
\label{sec:11}

Antiprotons are stored in a 300 MeV/c storage ring
equipped with 
cooling equipment and a Sibirian Snake. The circumference  of the ring
is chosen to be 150 m. There is no need for specially large acceptances
for the antiproton ring. 
The number of stored antiprotons,  $N_{\overline{p}}=10^{10}$, is assumed
to be evenly 
distributed over the circumference of the ring. The ring has a straight 
section in order to install an interaction zone
for the bremsstrahlung beam, see fig.7.
\begin{figure}{}
\resizebox{0.50\textwidth}{!}{%
  \includegraphics{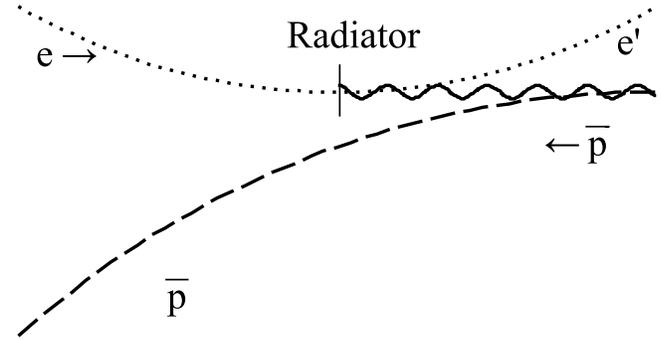}
}
\caption{Schematic picture of the interaction zone between
 the bremsstrahl beam and the antiproton beam. The interaction
 takes place in the vertical direction.
}
\label{fig:7}       
\end{figure}
The electron and antiproton beam enter into a strong magnetic field B (e.g.
3 T), perpendicular to their trajectories.
The electron beam is fed in the vertical direction from above 
approaching the trajectory of the antiproton beam, see fig.7.
Thereby, the electron beam
gets bent upwards in vertical direction, the antiprotons are bent
downwards.
A radiator is inserted into the electron beam line
at the point where the electron beam touches the field free trajectory of the
antiproton beam.  
After the radiator the electron beam will be bent upwards towards a beam dump. The
antiproton beam gets hit  at the entrance into the magnetic field by the bremsstrahl beam. 
The beam diameter of the antiproton beam is chosen to be 50 $%
\mu m.$
In order to get a large overlap of the antiproton and $\gamma $ -beam
a focus of around 30 $\mu m$ in diameter  of the electron beam on the
radiator should be accomplished. The beam entrance of the antiproton beam into the
magnetic field  and the end of the radiator are 10 mm apart. The distance of the
electron beam to that entrance point adds up to 230 $\mu m.$ The antiproton beam passes
the radiator at a distance of around 150 $\mu m.$ The characteristic bremsstrahl angle,
without multiple scattering,
leads to a distance of 25 $\mu m$ at the entrance point of the antiprotons
into the magnetic field.
That whole scheme is not optimized at all. It should demonstrate that it is
possible to prepare a bremsstrahl beam and keeping the electron and antiproton beam
separate. \textit{E.g.} by increasing the the magnetic field by a factor of two
and decreasing the momentum of the antiprotons by a factor of two the 
critical distances mentioned above reach the scale of 0.5 mm.

\subsection{The number of photons and the luminosity}
\label{12}

$N_{\gamma }$, the number of  polarised photons hitting the antiprotons,
is the last number missing in order to calculate the luminosity.
The relation
\begin{equation}
 \frac{dN_{\gamma }}{dE_{\gamma }^{\overline{p}-system}}=N_{e}%
\frac{\Lambda }{E_{\gamma }^{\overline{p}-system}}
\end{equation}
 with $\Lambda$ as the
fraction of the radiation length of the radiator, allows to calculate %
$N_{\gamma }$ with an adequate
accuracy:
$N_{\gamma}$=2.8 $\cdot 10^{13}s^{-1}$
with $\Lambda$ =0.06 and $N_{e}=1.2\cdot 10^{15}$ s$^{-1}$, coresponding to
a 200 $\mu$A beam current.
The integration range $155\preceq E_{\gamma }/MeV\preceq 230$ covers
the region of $\sigma _{1/2}$ dominance of the total absorption cross section.
The luminosity can be written as:
\begin{equation}
L=\frac{N_{\overline{p}}\cdot N_{\gamma }}{4\pi \cdot \sigma _{\overline{p}%
}\cdot \sigma _{\gamma }}\cdot \nu _{b} (cm^{-2}/s)
\end{equation}
where $\sigma _{\overline{p}}$ and $\sigma _{\gamma }$ characterize the
Gaussian transverse beam profiles. The diameters of the beams have been
chosen, as
discussed above, to d=50 $\mu m,$ and the frequency of the bunch crossing is
determined by the length of the antiproton storage ring and
the velocity of the antiprotons. With those numbers the luminosity adds up
to:
$L=\allowbreak 4.\,\allowbreak 3\times 10^{33}$ $%
cm^{-2}s^{-1}$.
That result is almost identical with the result of eq. 3. 
The uncertainties
entering as discussed above, mainly due to the chosen geometry in the
interaction zone, may add up to a factor of three to five.
The value of $\Lambda =0.06$ has been used routinely at Jefferson Lab.
for carrying out photon induced reactions.
An explicit calculation with the bremsstrahl spectrum, the degree of
polarisation of electron beam of 85\%, the unpolarized
contribution due to the not complete helicity transfer, see fig.4, leads to a
reduction of the figure-of-merit (FOM). That reduction depends
on which photon energy range will be covered. Selecting the threshold region, $%
E_{threshold}\prec E_{\gamma }/MeV\preceq 185,$ the figure-of-merit reaches 0.18.
Choosing, $E_{threshold}\prec E_{\gamma }/MeV\preceq 230,$ the region where $%
\sigma _{1/2}\succeq \sigma _{3/2},$ FOM adds up to 0.075.

\section{Preparation of an antineutron beam} 
\label{13}

For the $\gamma$-energy region $E_{threshold}\prec E_{\gamma }^{\overline{p}%
-system}/MeV\preceq 230$ the range of transverse momenta of the
photoproduced antineutrons covers a momentum range of $0\prec p_{\overline{n}%
}^{transverse}$ $MeV/c\preceq 125$ independently of the momenta of the
antiprotons. Accordingly, it is much more favourable to prepare an
antineutron beam using high energy storage rings for the antiprotons due to
kinematical focusing. However, looking at fig.7 it might  not be possible to
reach high energies because of keeping the electron and 
antiproton beam separate. For the case of a 10 GeV/c antiprotons,
certainly, the antiproton beam stays almost on his field free trajectory.
That means the radiator has to be taken out of the free field position and
positioned above it. The very low electron energy of only 10 MeV may still
allow to
find a configuration that  crossing of the bremsstrahlung with the
antiprotons becomes possible without interference of the electron and and
antiproton beam. In such a case a preparation of an antineutron beam with
high polarisation can be prepared. The intensities would be limited by the
production rate of antiprotons.

\section{Summary} 
\label{14}

In sect. 2 a method has been identified that could have the potential to
prepare a polarised antiproton beam in a storage ring. By making
use of two largely different photon absorption cross sections, $\sigma
_{1/2} $ and $\sigma _{3/2}$, one of the antiproton spin
projection gets enriched during the absorption processes and thus leads to a
polarised antiproton beam. A suitable circularly polarised photon beam can be prepared
by using a bremsstrahlung beam.
\newline
In addition, that method allows to extract from the stored antiprotons in
the ring, via the reaction $\overrightarrow{\gamma }+\overrightarrow{%
\overline{p}}\rightarrow \pi ^{-}+\overrightarrow{\overline{n}}$ a beam of
polarised antineutrons.
\newline
That method allows also, by switching the polarisation of the electron,
a measurement of the polarisation of
the stored antiproton beam to a high degree of precision by detecting
the produced antineutrons.
\newline
The advantage to use real photons resides in the fact
that one needs no large acceptance antiproton ring for the
antiprotons. At this point it is not clear,
how far in energy of the antiprotons the method is applicable.
At an antiproton energy of 10 GeV/c \textit{e.g.} the trajectory
of the antiproton is almost a straight line. However, the electron
energy decreases down to 10 MeV, which, possibly, allows a very close
side way distance of the radiator to the antiproton beam.

\end{document}